\documentclass[aps,twocolumn,showpacs,pra,longbibliography,superscriptaddress]{revtex4-2}
\usepackage{array}\usepackage{calc}

\usepackage{mathtools}
\usepackage{bm}
\usepackage{dsfont,amsthm,amsbsy}
\usepackage{verbatim}
\usepackage{amssymb}
\usepackage{amsmath}
\usepackage{bbm}
\usepackage{graphicx}
\usepackage{epstopdf}
\usepackage{subfigure}
\usepackage{natbib}
\usepackage{epsfig}
\usepackage{amsfonts}
\usepackage{mathrsfs}
\usepackage{sidecap}
\usepackage{lipsum}
\usepackage[toc,page,title,titletoc,header]{appendix}
\usepackage[colorlinks,linkcolor=blue,citecolor=blue,anchorcolor=blue,urlcolor=blue]{hyperref}
\usepackage{hyperref}
\usepackage{resizegather}
\usepackage{tikz}
\usepackage{float}
\usepackage{mathbbol}
\usepackage[normalem]{ulem}
\usepackage{cancel}
\usepackage{upgreek}

\usepackage[colorlinks,linkcolor=blue,citecolor=blue,anchorcolor=blue,urlcolor=blue]{hyperref}
\usepackage{hyperref}

\begin{document}
\title{Ballistic Entanglement Cloud after a Boundary Quench}

\author{Bedoor Alkurtass}
\affiliation{Department of Physics, College of Science, Sabah Al Salem University City, Kuwait University, P.O. Box 2544, Safat 1320, Kuwait}
\author{Abolfazl Bayat}
\affiliation{Institute of Fundamental and Frontier Sciences, University of Electronic Science and Technology of China, Chengdu 611731, China}
\affiliation{Key Laboratory of Quantum Physics and Photonic Quantum Information, Ministry of Education, University of Electronic Science and Technology of China, Chengdu 611731, China}

\author{Pasquale Sodano}
\affiliation{I.N.F.N., Sezione di Perugia, Via A. Pascoli, I-06123, Perugia, Italy} 

\author{Sougato Bose}
\affiliation{Department of Physics and Astronomy, University College London, Gower Street, London WC1E 6BT, UK}

\author{Henrik Johannesson}
\affiliation{Department of Physics, University of Gothenburg, SE 412 96 Gothenburg, Sweden}

\begin{abstract}
Entanglement has been extensively used to characterize the structure of strongly correlated many-body systems. Most of these analyses focus on either spatial properties of entanglement or its temporal behavior. 
Negativity, as an entanglement measure, quantifies entanglement between different non-complementary blocks of a many-body system. Here, we consider a combined spatial-temporal analysis of entanglement negativity in a strongly correlated many-body system to characterize complex formation of correlations through  non-equilibrium dynamics of such systems. A bond defect is introduced through a local quench at one of the boundaries of a uniform Heisenberg spin chain. Using negativity and entanglement entropy,  computed by the time-dependent density matrix renormalization group, we analyze the extension of entanglement in the model as a function of time. We find that an entanglement cloud is formed, detached from the boundary spin and composed of spins with which it is highly entangled. The cloud travels ballistically in the chain until it reaches the other end where it reflects back and the cycle repeats. The revival dynamics exhibits an intriguing contraction (expansion) of the cloud as it moves away from (towards) the boundary spin.
\end{abstract}
\date{\today}

\maketitle

\section{Introduction}

Quantum entanglement, as one the most striking features of quantum mechanics, has been identified as a resource which allows for outperforming classical systems in communication~\cite{bennett1993teleporting,bennett1992communication}, computation~\cite{briegel2001persistent} and sensing~\cite{giovannetti2004quantum}. In the last two decades, a lot of efforts have been dedicated to quantification of entanglement~\cite{plenio2005introduction,horodecki2009quantum}, its generation~\cite{monz201114,gong2019genuine,wang2016experimental,blatt2008entangled}, detection~\cite{guhne2009entanglement} and application~\cite{erhard2020advances}. Strongly correlated many-body systems naturally support a significant amount of entanglement in their structure~\cite{amico2008entanglement,bayat2022entanglement}.  Entanglement in such systems have been mainly investigated in two different directions, namely characterizing its (i) spatial properties or (ii) temporal features. In (i), entanglement between different blocks of a many-body system is investigated in terms of the block-sizes and their spatial distance~\cite{calabrese2009entanglement_entropy,calabrese2009entanglement,calabrese2011entanglement,wichterich2009scaling,sorensen2007quantum,bayat2010negativity,bayat2012entanglement,alkurtass2016entanglement,roy2022entanglement}. Such studies reveal that while the eigenstates at the edge of the spectrum, i.e. low- and high-energy eigenstates, support area-law entanglement, the mid-spectrum eigenstates provide volume-law entanglement~\cite{eisert2010colloquium}.  In (ii), temporal behavior of entanglement is  typically studied in the non-equilibrium dynamics of a many-body system induced by a quantum quench~\cite{calabrese2005evolution,calabrese2011quantumquench,bayat2010entanglement,de2006entanglement}. It has been shown that entanglement may indeed be generated through the dynamics and can linearly grow from zero to area-law and then eventually saturate to a volume-law value~\cite{calabrese2016quantum_cardy,fagotti2008evolution,kim_ballistic_2013,de2006entanglement,altman2015universal,bardarson2012unbounded}. Despite all these studies,  the combined spatial {\em and} temporal behavior of entanglement negativity in a many-body system has not received much attention.

In this article, we try to make up for this by presenting an analysis of the spatial-temporal structure of entanglement propagation after a quantum quench in a bounded many-particle system. The more general problem of entanglement spreading following a quantum quench has attracted considerable interest in recent years, much spurred by the desire to understand equilibration of isolated many-body systems \cite{Polkovnikov2011,Gogolin2016,calabrese2016quantum_cardy,Essler2016,Mitra2018,Abanin2019}. The underlying motivation is experimental, where advances in quantum technologies have enabled the observation of unitary time evolution in various physical platforms, including 
ultra-cold atoms~\cite{Lewenstein2007,Altman2016},  ion traps~\cite{blatt2012quantum,zhang2017observation,zhang2017observation2}, and superconducting quantum simulators~\cite{neill2016ergodic, mi2022time,roushan2017spectroscopic,guo2021stark,guo2021observation,gong2021quantum, gong2021experimental}. The case of a {\em local} quench $-$  where the Hamiltonian that governs the evolution is suddenly changed in a local region of space $-$ has mainly been studied in one-dimension (1D) employing the ``cut and glue" protocol suggested by Calabrese and Cardy \cite{Calabrese2007}. In that work, a uniform system is cut in two pieces, each prepared in its respective ground state. The two pieces are then glued together by a local quench and the time evolution of the entanglement for a chosen bipartition of the system is being studied. The early results from conformal field  theory (CFT) suggested that the spreading of entanglement originates from pairs of freely propagating entangled quasiparticles produced by the quench at the gluing point \cite{Calabrese2007}. This picture, as well as specifics of the CFT results $-$ later generalized to finite systems \cite{Stephan2011} and disjoint partitions \cite{Asplund2014,Wen2015} $ -$ has been supported by several analytical and numerical studies of 1D lattice models \cite{Gobert2005,Eisler2007,Eisler2007b,Kleine2008,Hsu2009,Igloi2009,Eisler2012,Collura2013,Eisler2014, dell2019dynamics}. 

Here, using a different type of protocol, we explore the spatial-temporal behavior of entanglement after a local quench on a uniform and finite 1D lattice system, with the quench creating a bond defect at one of the boundaries of the lattice. By this, the interaction of a boundary degree of freedom with the bulk gets modified and becomes different from the uniform interaction in the bulk. With an eye towards a proposal to use spin chains as quantum communication channels, with entanglement spreading being a key feature \cite{Bose2003}, we take the system to be a spin-1/2 Heisenberg chain with open boundary conditions. With this, the boundary degree of freedom is simply a spin at one of the endpoints of the chain.

Employing negativity \cite{lee2000partial,vidal_computable_2002} as a measure of entanglement and using the time-dependent renormalization group (tDMRG) \cite{schollwock2005density} allows us to track the entanglement of the boundary spin with the rest of the system in time as well as in space. The numerical results reveal that, just after the quench, a group of spins are strongly entangled with the boundary site, in the immediate vicinity of it. We call those spins an "entanglement cloud”. As the system evolves unitarily, the neighboring spins are significantly unentangled from the boundary, while more distant spins are progressively entangled. This result can be interpreted as the cloud travelling ballistically in the chain and being detached from the boundary spin. The ballistic propagation of the cloud before its first reflection at the other end of the lattice may heuristically be understood from a quasiparticle picture analogous to that of the cut and glue scenario. More intriguing is the long-time recurrent dynamics of the detached cloud. While still ballistic, the cloud now contracts (expands) as it moves away from (towards) the boundary spin. To substantiate this striking behavior we run an independent check by defining a boundary entanglement entropy and show, again using tDMRG, that its space-time pattern is in agreement with the results extracted from the negativity measure. 


The layout of the paper is as follows: In section~\ref{sec:model} we introduce the model and define an entanglement cloud, using negativity as entanglement measure. In section~\ref{sec:negativity} we describe a local quench in a Heisenberg Hamiltonian which 
results in a complex entanglement structure in the system. 
Through a comprehensive analysis of spatial and temporal behavior of entanglement negativity, we provide a heuristic picture of a ballistically travelling entanglement cloud which moves along the chain and gets reflected after hitting the boundaries.  In section \ref{sec:entropy}, the block entanglement entropy is studied and the contribution of the boundary spin to the entropy is analyzed. Finally we conclude with a summary in section~\ref{sec:summary} in which we also provide an outlook for further studies.

\section{Model and entanglement length} 
\label{sec:model}
We consider an antiferromagnetic spin-1/2 Heisenberg chain with open boundary conditions, and write its Hamiltonian as
\begin{equation}
\mathcal H(J^{\prime})= J^{\prime} \boldsymbol{S}_1\cdot\boldsymbol{S}_{2}
+ \sum_{l=2}^{N-1}\boldsymbol{S}_l\cdot\boldsymbol{S}_{l+1} 
\label{eq:QK_Hamiltonian}
\end{equation}
where $N$ is the length of the chain and where $\boldsymbol{S}_{l} = \boldsymbol{\sigma}_{l}/2$ are spin-1/2 operators acting on site $l$, with 
$\boldsymbol{\sigma}_{l} = (\sigma^x_l,\sigma^y_l,\sigma^z_l)$ the vector of Pauli matrices. We have here singled out the coupling $J^{\prime}\!>\!0$ between the leftmost boundary spin and its nearest neighbour $-$ to be used as a quench parameter $-$ with the rest of the chain supporting a nearest-neighbour coupling of unit magnitude.

\begin{figure}[htbp]
\centering
\includegraphics[width=0.45\textwidth]{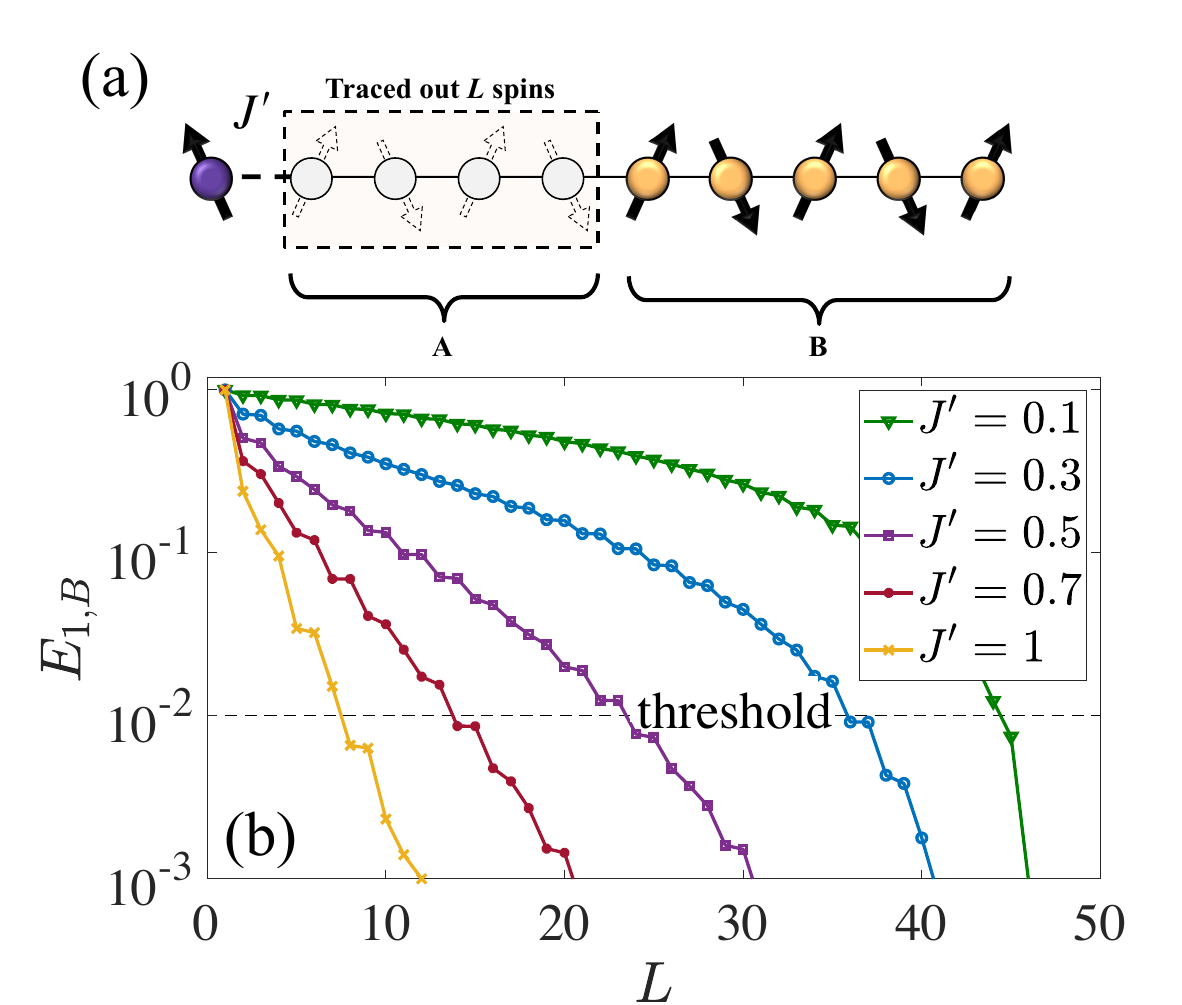}
   \caption{{\protect\footnotesize (a) Evaluation of the entanglement length $\xi$: $L$ spins in the region $A$ are traced out and the entanglement between the boundary spin and region $B$, as measured by the negativity $E_{1,B}$, is evaluated. (b) $E_{1,B}$ decreases exponentially with $L$ for the uniform chain with $J^{\prime}\!=1$. In the presence of a bond defect, here with magnitudes $J^{\prime}\!= 0.1, 0.3, 0.5, 0.7$, the decay of $E_{1,B}$ is slower, still roughly exponential but with a smaller decay constant. In all cases $\xi$ can be extracted by setting a threshold, here chosen as $10^{-2}$. The system size is taken to be $N=50$.}}
\label{QK_Fig1}
\end{figure}

For any given quantum state $|\Psi\rangle$, we define an entanglement length $\xi$ as the number of spins that are significantly entangled with the boundary spin. To evaluate $\xi$ we first divide the chain into three regions, shown in Fig.~\ref{QK_Fig1}(a), such that the boundary spin is the leftmost spin at site $\ell =1$, and where regions $A$ and $B$ are made up of $L$ and the remaining $N\!-\!L\!-\!1$ spins, respectively. For the purpose of computing the entanglement between the boundary spin and region $B$, we define the reduced density matrix $\rho_{1,B}=\mathrm{Tr}_A |\Psi\rangle \langle \Psi|$ by tracing out the spins in region $A$. As entanglement quantifier we use negativity \cite{lee2000partial,vidal_computable_2002}, defined by
\begin{equation} \label{eq:negativity}
E_{1,B}=\sum_{i}|a_i|-1,
\end{equation}
where $a_i$ are the eigenvalues of the partially transposed density matrix $\rho_{1,B}^{T_{B}}$. In Fig. 1(b) we display $E_{1,B}$ in
a semi-log plot, computed by DMRG for a few different values of $J'$, for the case that $|\Psi\rangle$ is the normalized ground state of the corresponding Hamiltonian in (1) with $N=50$. As seen in the figure, the entanglement for the uniform chain $(J^{\prime}\!=\!1)$ shows an exponential decay as $L$ increases, with the decay in the presence of a bond defect $(J^{\prime}\!<\!1)$ still being roughly exponential but with a smaller decay constant. 

By setting an entanglement threshold, we can identify the entanglement length $L=\xi$ beyond which the entanglement between the boundary spin and region $B$ is below the chosen threshold value. This implies that the boundary spin is nearly maximally entangled with region $A$ of size $L=\xi$. In Fig.~\ref{QK_Fig1}(b) we have marked $10^{-2}$ as threshold value and this is also the value we use throughout this paper. As we show below, the choice of threshold $10^{-2}$ is sufficiently small not to significantly impact our results. The DMRG results are obtained for chains of length $N=50$, implying a truncation error $\sim 10^{-7}$.

\section{Boundary quench dynamics: negativity probe} 
\label{sec:negativity}

We now prepare the system in the normalized ground state $|\Psi_0\rangle$ of the uniform Heisenberg chain, i.e., with $J^{\prime}\!=1$ in eq. (\ref{eq:QK_Hamiltonian}). In this state the boundary spin (i.e., site $l {=}1$) is most entangled with its nearest neighbor \cite{OConnor2001,Chen2006} and the entanglement length $\xi$  is small; cf. Fig. ~\ref{QK_Fig1}(b). At time $t=0$ the coupling to the boundary spin is suddenly quenched to $J^{\prime}\!< 1$ and the system starts to evolve as $|\Psi(t)\rangle =e^{-i\mathcal{H}(J')t} |\Psi_0(J'{=}1)\rangle$, with ${\cal H}(J')$ the quenched Hamiltonian in Eq. (\ref{eq:QK_Hamiltonian}). We use a small value of $J'$ such that the ground state $|\Psi_0(J')\rangle$ of ${\cal H}(J')$ supports a large entanglement length, comparable with the system size $N$. The overlap, call it $F$, between the quenched state $|\Psi(t)\rangle$ and the ground state of the quenched Hamiltonian $|\Psi_0(J')\rangle$ is time independent, and, by unitarity, is given by
\begin{eqnarray}
\nonumber F&=&|\langle \Psi_0(J')| \Psi(t)\rangle|^2\\
\nonumber &=& |\langle \Psi_0(J')| e^{-i\mathcal{H}(J')t}|\Psi_0(J'{=}1)\rangle|^2\\
&=&|\langle \Psi_0(J')| \Psi_0(J'{=}1)\rangle|^2
\end{eqnarray}

Since the quench is local and only affects the coupling between two single spins, $F$ approaches unity for sufficiently large $N$. Unlike $F$, the entanglement length $\xi(t)$ defined as above, and computed for the state $|\Psi(t)\rangle$, is time dependent and may grow to large values.

\begin{figure}[htbp]
\centering
\includegraphics[width=0.48\textwidth]{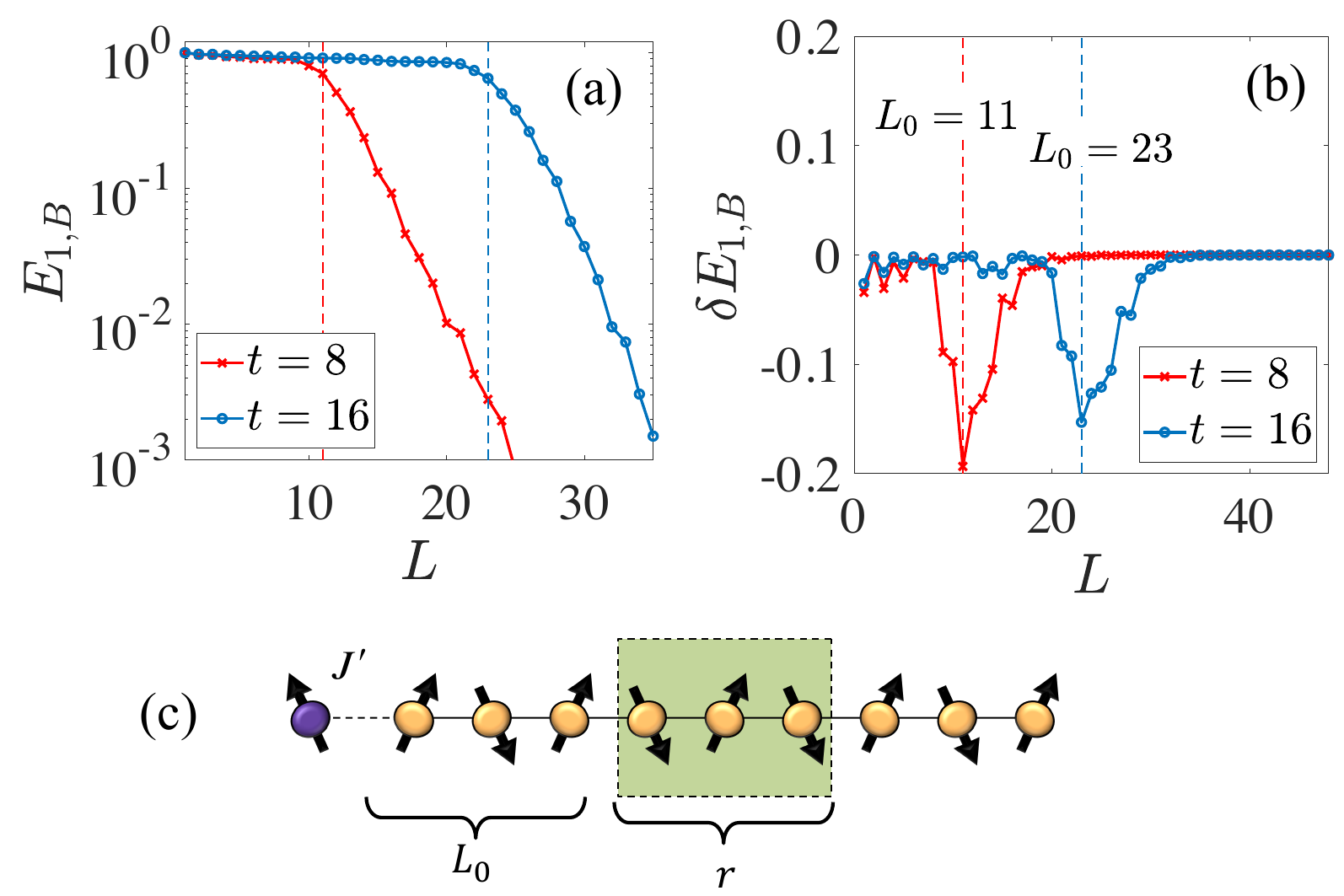}
   \caption{{\protect\footnotesize (a) Entanglement $E_{1,B}$ as a function of $L$ for $N=50$ at two instants of time after a quench $J^{\prime}\!= 1 \rightarrow J^{\prime}\!= 0.1$ at $t=0$. $E_{1,B}$ decreases very slowly up to $L=L_0$, then starts decreasing faster. The vertical line separates the two regions. (b) The difference $\delta E_{1,B}$ as a function of $L$ is used to extract $L_0$. (c) Schematic snapshot of the system after the quench.}}
\label{QK_Fig2}
\end{figure}

To map out $\xi(t)$, we compute the negativity $E_{1,B}(L,t)$ between the boundary spin and region $B$ at different instants of time $t$, with $B$ separated from the boundary spin by a distance $L$. While $E_{1,B}$ decreases roughly exponentially with $L$ in the ground state $|\Psi_0(J')\rangle$, see Fig.~\ref{QK_Fig1}(b), a distinct behaviour emerges in the nonequilibrium postquench states $|\Psi(t)\rangle$. As an illustration, in Fig.~\ref{QK_Fig2}(a) we plot $E_{1,B}$ as a function of $L$ at two different times $t=8$ and $t=16$ with $J'=0.1$. The semi-log plot exhibits two different regions for each of the two times: In the first region, characterized by a weakly tilted plateau in the figure, $E_{1,B}$ decays very slowly with an increase of $L$ up to $L=L_0(t)$. This implies that the boundary spin is {\em weakly} entangled with the spins in the region with $L<L_0(t)$. In the second region, $L>L_0(t)$, $E_{1,B}$ exhibits a fast decay, similar to that in the ground state, and most of the entanglement of the boundary spin is with the spins in this region. Given this observation, one may think of the spins in the second region as forming a {\em cloud of strong entanglement} with the boundary spin.
Although the two distinct regions are clearly noticed from Fig.~\ref{QK_Fig2}(a), they are not sharply separated. To find the value of $L_0(t)$ that separates the two regions, the difference
\begin{equation}
\delta E_{1,B}(L) = E_{1,B}(L)-E_{1,B}(L-1)
\end{equation}
is evaluated for each instant of time. Then $L_0(t)$ is defined as the value of $L$ at which $\delta E_{1,B}$ is at a minimum, as shown in Fig.~\ref{QK_Fig2}(b). 

Since the transition between the two regions is gradual, sometimes it is difficult to extract a precise value of $L_0$ as the minimum of $\delta E_{1,B}$ extends over a few neighboring sites. 
Nonetheless, we expect that in such a case the maximum sharpens into a peak as the system size is increased, and the estimate of $L_0(t)$ becomes more precise. For any given instant $t$, the entanglement length $\xi(t)$ is found by computing the length $L$ at which $E_{1,B}(L,t)$ reaches the chosen entanglement threshold $10^{-2}$. The instantaneous range of the entanglement cloud, call it $r(t)$, is then given by $r(t) = \xi(t)-L_0(t)$; cf. Fig.~\ref{QK_Fig2}(c). Referring to Fig.~\ref{QK_Fig2}(a), note that the choice of the entanglement threshold does not matter much since it intersects with the curve of $E_{1,B}$ at $L>L_0$, where the entanglement decays exponentially by a further increase of $L$.

\begin{figure}[hbtp]
\centering
\includegraphics[width=0.5\textwidth]{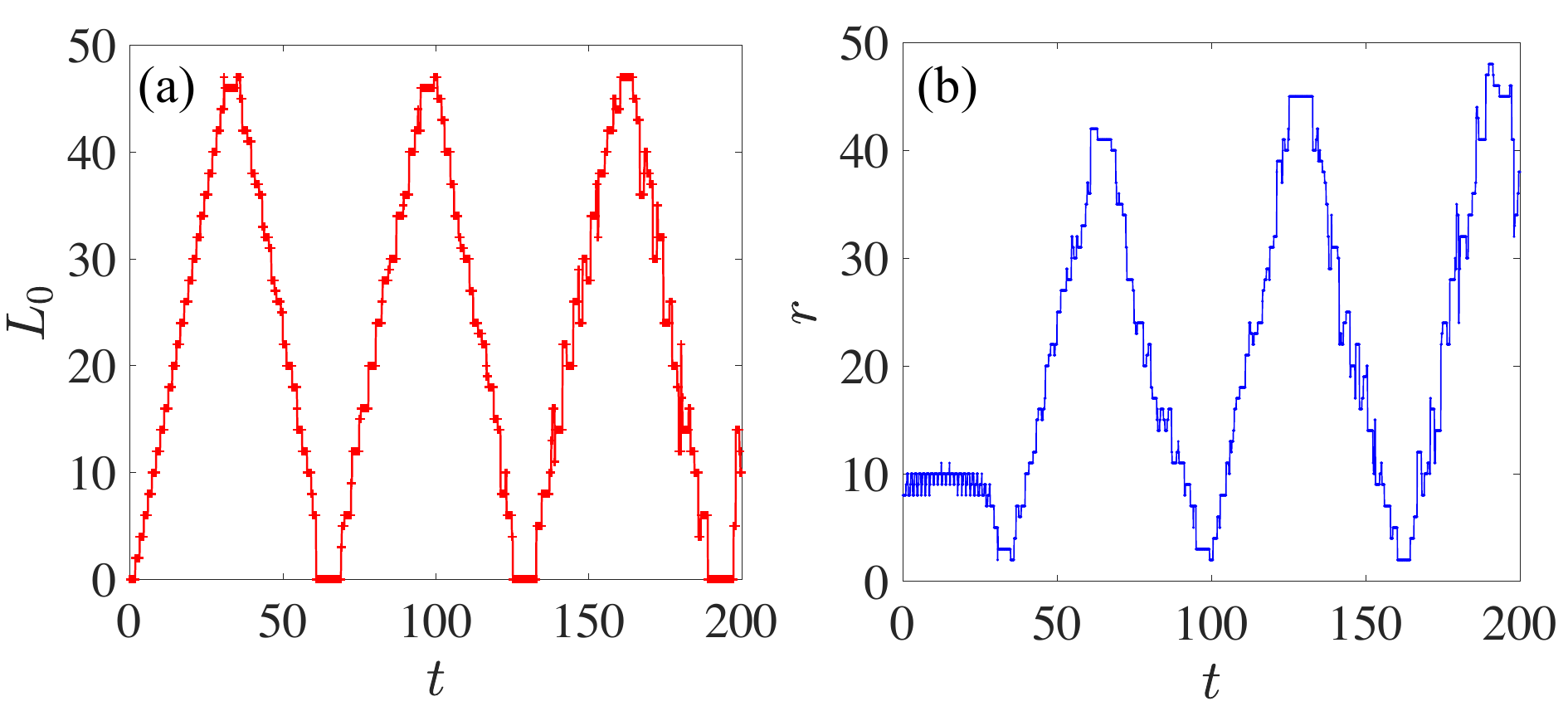}
\includegraphics[width=0.4\textwidth]{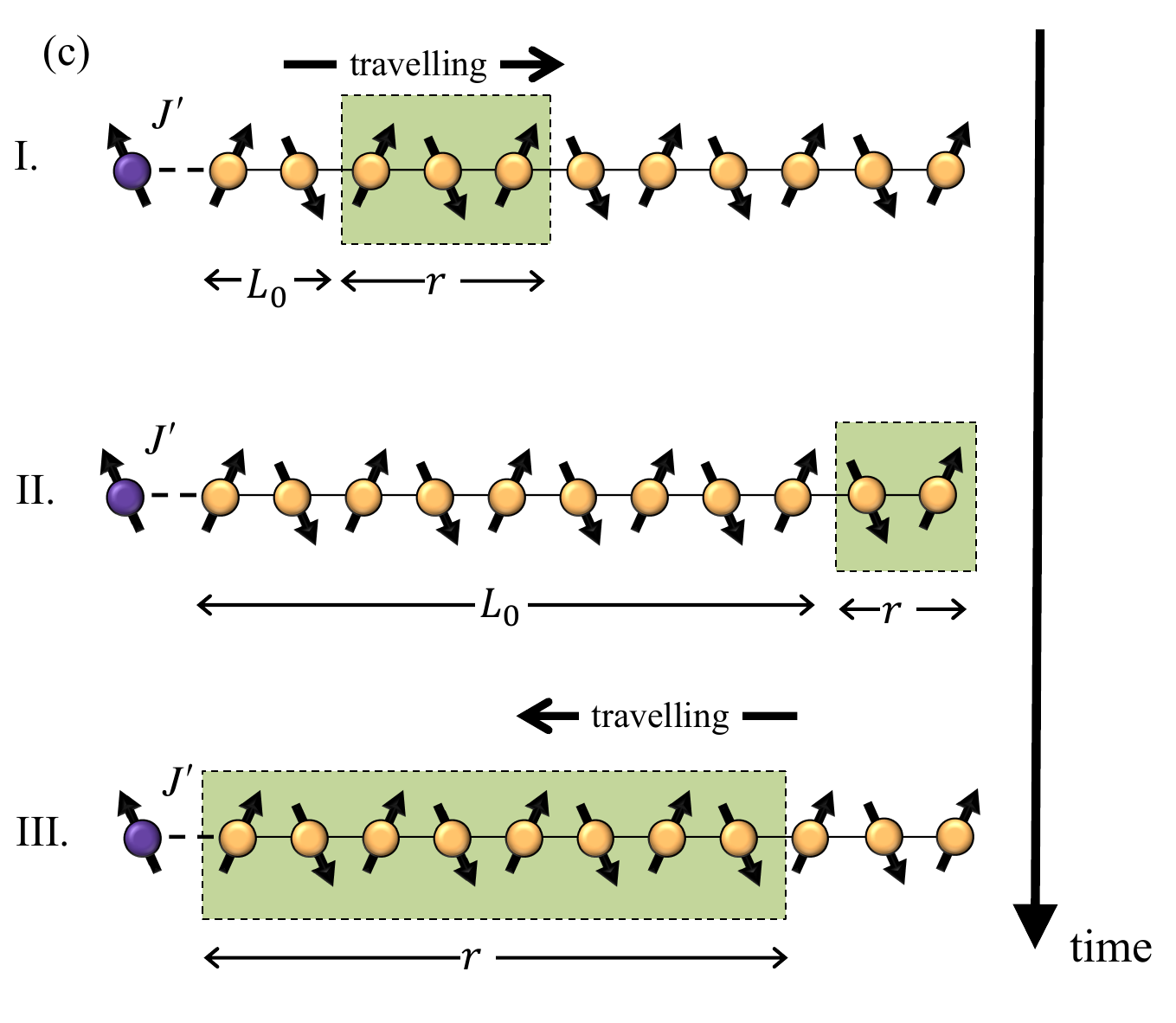}
   \caption{{\protect\footnotesize (a) Time evolution of the distance $L_0$ between boundary spin and entanglement cloud. (b) Range $r$ of the entanglement cloud after a quench $J'=1 \rightarrow J'=0.1$ at $t=0$ for a chain of length $N=50$. (c) Schematic snapshots of the time-evolved entanglement cloud after the quench.}}
\label{travelling}
\end{figure}

Fig.~\ref{travelling}(a) and (b), respectively, display the behavior of $L_0(t)$ (cloud distance from the boundary spin) and $r(t)$ (cloud range). It is seen that $L_0(t)$ oscillates with time which implies that the entanglement cloud is travelling back and forth in the chain. Interestingly, $r(t)$ remains almost constant initially (up to small fluctuations) until the entanglement cloud approaches the other end of the chain when it tapers off to eventually cover only a few sites at a time $t\approx 30$.  In other words, the leftmost boundary spin now effectively gets entangled only with the last few spins close to the right endpoint. At this instant,  the extent of the plateau reaches its largest value, $L_0 \approx N-1$, cf. the first peak of $L_0$ in Fig.~\ref{travelling}(a). After that, $L_0(t)$ decreases while $r(t)$ increases, indicating that the cloud expands as it moves back towards the boundary spin after having been reflected at the right endpoint of the chain. Eventually, at its maximal extension, the cloud hits the boundary spin at site $l {=} 1$ at a time $t\approx 65$, gets reflected and shrinks again. The cycle of propagation and reflection of the entanglement cloud then repeats, with the maximal size of the cloud growing slowly with each revival; cf. Fig.~\ref{travelling}(b). The amount of entanglement between the boundary spin and the cloud, read off at the endpoint of a plateau as in Fig.~\ref{QK_Fig2}(a), varies with time. During the first cycle, it varies between  $0.99$ (given the entanglement threshold of $0.01$) and $0.6 \pm 0.05$. For later cycles, the uncertainties in the numerical values of $L_0(t)$ (that locates the instantaneous endpoint of a plateau), become somewhat larger, implying a slightly less precise estimate of the lower bound $0.6\pm 0.05$. The values of $L_0(t)$ during the first cycle is estimated up to $\pm 1$ site, while for later cycles the error becomes $\pm 3$ sites. The late-time larger uncertainties in $L_0(t)$ are expected to be due to small instabilities that build up with time in this type of tDMRG computation, showing up also in the scatter plots for $L_0(t)$ and $r(t)$ in Fig.~\ref{travelling}.

The magnitude of the slope of the red $L_0$-graph, Fig.~\ref{travelling}(a), is constant throughout the cycles and show that the speed of propagation of the entanglement cloud is ballistic.  Its value $\approx 1.51$ (in units where the lattice spacing is fixed to unity and with $\hbar = 1$), is close to the Lieb-Robinson bound \cite{lieb_finite_1972}, equal to $\pi/2$ when applied to a spin-1/2 antiferromagnetic Heisenberg chain with unit bulk coupling \cite{Vlijm2016,Gruber2021}. In Fig.~\ref{travelling}(c) we have schematically illustrated a few snapshots of the first cloud cycle.  

The initial formation and propagation of the entanglement cloud invites an interpretation somewhat similar to that of the quasiparticle picture drawn from the cut and glue scenario \cite{Calabrese2007}: The boundary spin gets entangled with a propagating quasiparticle induced by the local quench, here tentatively taken as a wave packet built out of the spinons in the excitation spectrum of the Heisenberg chain \cite{Faddeev1981}. The width of the wave packet may be envisaged to roughly correlate with the range of the propagating entanglement cloud, being approximately constant right after the quench; cf. Fig.~\ref{travelling}(b). As the quasiparticle moves away from the bond defect it leaks some entanglement into its surrounding, leaving behind a dilute trail of entanglement with the boundary spin $-$ signaled by the weak tilt of the plateaux; cf. Fig.~\ref{QK_Fig2} (a). Different from generic applications of the cut and glue protocol, the initial prequench ground state has a lower energy $(E^0_{J^{\prime}=1}=-22.0)$ than the ground state of the quenched Hamiltonian $(E^0_{J^{\prime}=0.1}=-21.5)$ (with numbers obtained from DMRG with $N=50$). It follows that the conjectured quasiparticle should not be thought of as an excitation drawn from a preexisting energy reservoir; instead its energy is supplied by the quench and thus lives on top of the prequench ground state. Moreover, the nonconservation of momentum in the quench process is also different from the standard cut and glue scenario \cite{Calabrese2007} as the quench in our case takes place in the presence of open boundaries. Since spinons are produced in pairs \cite{Faddeev1981}, our conjectured quasiparticle picture presupposes that half of the spinons get trapped at the bond of the local quench, effectively manifested as an excitation of the boundary spin. Let us stress that a confirmation of our conjecture must await further advances in theory; there are as yet no exact or rigorous results on spinon states bound to an open boundary with a defect.

While only provisory in character, the scenario just outlined checks with the propagation speed of the entanglement cloud, being close to the Lieb-Robinson bound as set by the maximal spinon velocity $v=\pi/2$ in a spin-1/2 Heisenberg chain \cite{Vlijm2016}.  It is also worth emphasizing that as a result of the non-equilibrium dynamics, the quantum state of the system becomes very complex. In such complex state, entanglement between the boundary spin and the bulk becomes multipartite, which may not be reflected in entanglement between the boundary and other individual spins in the bulk. In particular, the bipartite entanglement between the boundary spin and all other spins is found to be short-ranged and does not capture the cloud.

Finally, we would like to mention that one may think of considering the mutual information as a probe of the correlations as it quantifies both classical and quantum correlations, in contrast to the negativity which only captures quantum correlations. However, we found that using the mutual information to define a highly correlated cloud gave the same physical picture of a ballistically moving cloud. It would be interesting to classify the correlations into classical and quantum \cite{Henderson2001classical,Modi2010classical}, however, such analysis can be NP-hard \cite{Ioannou2007complexity,Huang2014discord} and beyond the scope of the current work.

\section{Boundary quench dynamics: entanglement entropy} 
\label{sec:entropy}

The revival dynamics of the entanglement cloud $-$ with repeating cycles of reflection, expansion, and contraction, cf. Fig.~\ref{travelling}(c)  $-$ is difficult to conceptualize already at a heuristic level. Independent results are therefore critical so as to support and complement the picture drawn from the negativity probe. For this reason we have performed an independent numerical analysis, looking for imprints of the entanglement cloud on the block entanglement entropy.

\begin{figure}[hbtp]
\centering
\includegraphics[width=0.5\textwidth]{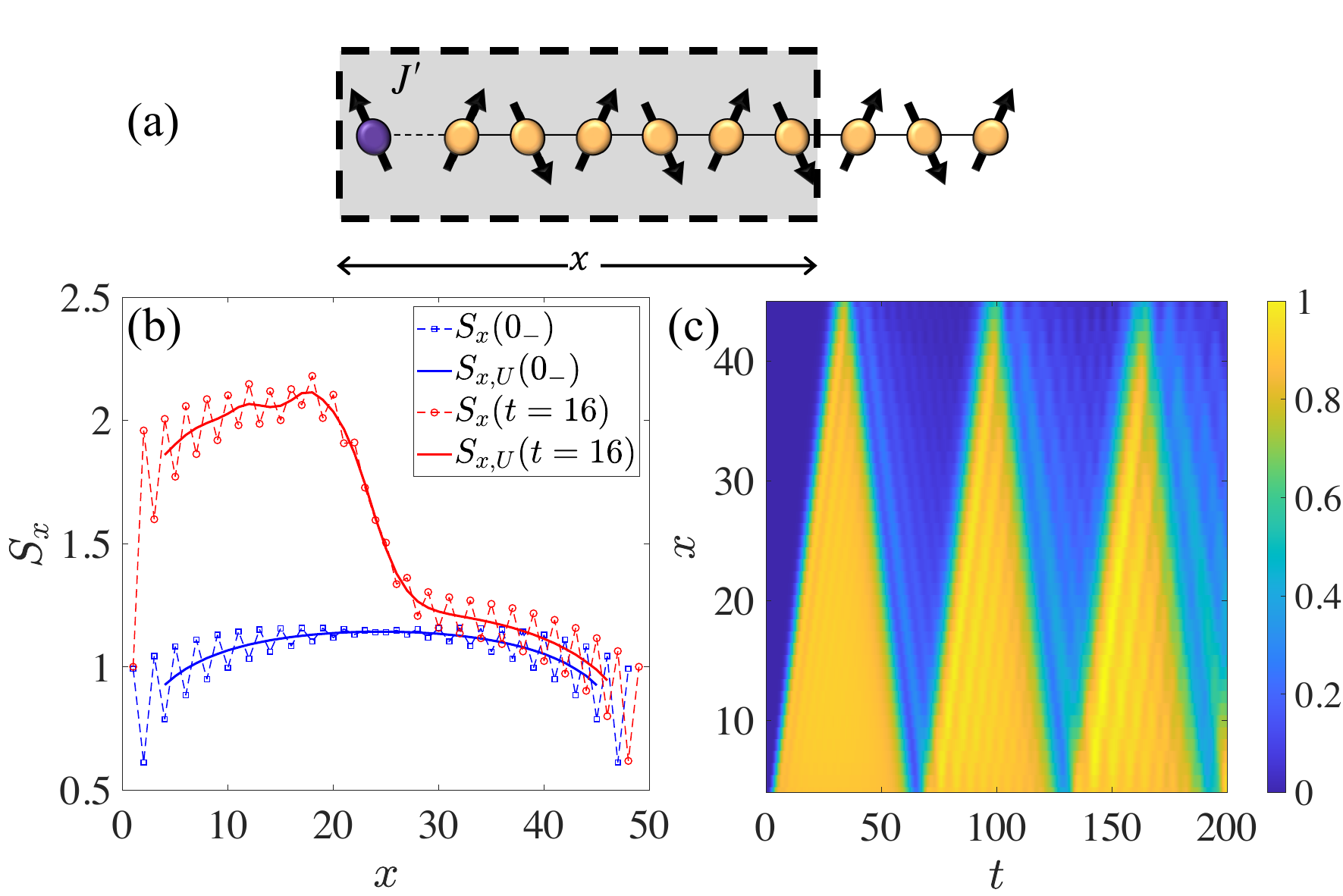}
   \caption{{\protect\footnotesize (a) Schematic figure of the block used to compute the entanglement entropy $S_x(t)$. (b) The entanglement entropy $S_x(t)$ before (blue) and at time $t\!=\!4$ after (red) the quench $J'\!=\!1 \rightarrow J'\!=\!0.1$ at $t\!=\!0$ for a chain of length $N\!=\!50$. Uniform [oscillating] parts are represented by solid [dashed] lines. (c) Time evolution of the boundary spin contribution $S_{bs}$ after the quench.}}
\label{Entropy91}
\end{figure}

To study this, we construct the reduced density matrix $\rho_x(t) = \mbox{Tr}_{x} |\Psi(t)\rangle \langle \Psi(t) |$ of a block of length $x$ (including the boundary spin,  $l = 1$,  cf. Fig.~\ref{Entropy91}(a)), tracing out all spins outside the block as indicated by $x$, and with  $|\Psi(t)\rangle$ being the postquench state at time $t$. The von Neumann entropy 
$S_x(t) = -\mbox{Tr}[\rho_x(t)\log_2 \rho_x(t)]$ quantifies the instantaneous entanglement between the block and the rest of the system \cite{Bennett1996}. With the block starting at one end of the chain, the boundary is known to cause an additional oscillating term in the entanglement entropy that decays with the distance from the boundary \cite{sorensen2007quantum, laflorencie2006boundary}, in our case $-$ off equilibrium $-$ depicted by the dashed lines in Fig.~\ref{Entropy91}(b). The uniform part of the entanglement entropy, call it $S_{x,U}$, can be extracted numerically by the same procedure as for the ground state \cite{sorensen2007quantum}, with the result
represented by the solid lines in Fig.~\ref{Entropy91}(b).

The contribution to the uniform entanglement entropy from the boundary spin in the presence of the bond defect introduced by the quench at time $t\!=\!0$ can be defined as
\begin{equation}
S_{bs} = S_{x,U}(t)-S_{x,U}(0_{-}), \ \ t>0,
\end{equation}
where $S_{x,U}(0_{-})$ [$S_{x,U}(t)$] is the uniform part of the block entropy before [after] the quench. Pictorially, $S_{bs}$ represents the difference between the two solid lines in Fig.~\ref{Entropy91}(b).

Again using tDMRG on a chain of length $N\!=\!50$, now for computing $S_{bs}$, we obtain the data displayed in Fig.~\ref{Entropy91}(c). To understand the oscillations of $S_{bs}$ with time as seen in the figure, let us focus on a block of length $x=25$, i.e., the postquench entanglement entropy of the first half of the chain including the boundary spin. Initially, right after the quench, $S_{bs}$ is small, consistent with the expected short range $r$ of the entanglement cloud, as seen from the negativity probe in Fig.~\ref{travelling}(b). Around time $t\approx 20$, $S_{bs}$ exhibits a rapid increase. This can be understood from Fig.~\ref{travelling}(a): around the same time, $t \approx 20$, the entanglement cloud crosses the block boundary $(L_0 \approx 25)$ after which the boundary spin quickly becomes entangled with spins outside the block. Going back to Fig.~\ref{Entropy91}(c), when $t \approx 48$, $S_{bs}$ is seen to suffer a fast drop. This is also in agreement with Fig.~\ref{travelling}(a) and (b) which shows that around this time the entanglement cloud again crosses the block boundary but in the opposite direction, having been reflected at the other end of the chain. As a result, the boundary spin gets entangled primarily with spins within the block, causing the fast drop of $S_{bs}$. By similarly following the oscillations in Fig.~\ref{Entropy91}(c) for later times and for any choice of block length $x$, one finds a good match with the negativity results in Fig.~\ref{travelling}(a) and (b).

\section{Summary and outlook} 
\label{sec:summary}

Summing up our results: Through a comprehensive spatial-temporal characterization of entanglement, we have shown that in a spin-1/2 antiferromagnetic Heisenberg chain, a local quantum quench that weakens a single boundary bond  produces a ballistic entanglement cloud comprised of spins in the bulk that are highly entangled with the spin at the boundary. After the rapid formation of the cloud right after the quench, it contracts when approaching the other end of the chain, then gets reflected and expands as it moves back towards the boundary spin and again gets reflected. This cycle repeats, with a small increase of the maximal size of the cloud with each cycle. 

In parallel with searching for an explanation of the intriguing cloud revivals, it should be interesting to investigate analogies with nonequilibrium Kondo physics \cite{lobaskin2005crossover, heyl2010interaction, ratiani2010nonequilibrium, latta2011quantum, joho2012transient, schiro2012nonequilibrium, munder2012anderson, medvedyeva2013spatiotemporal, vasseur2013crossover, lechtenberg2014spatial,  kleine2014real, kennes2014universal,  vinkler2014thermal,  schiro2014transient, ghosh2015dynamics, bragancca2021quench, stocker2022entanglement, bellomia2024quasilocal, aikebaier2024machine}. As is well known, in equilibrium and at low temperatures, a localized impurity in a fermionic host gets entangled with, and thereby screened by, the spins of the fermions \cite{Hewson, sorensen2007quantum, v2020observation}. At zero temperature the entanglement becomes maximal, leading to perfect screening of the impurity spin \cite{yosida}. The buildup of the screening cloud after a local quench has been studied for a special value of the impurity coupling that allows for a free-particle representation (Toulose limit) \cite{medvedyeva2013spatiotemporal}. A complementary approach is suggested by our work, using that a spin-chain version of the Kondo model can be obtained by adding a fine-tuned next-nearest neighbor coupling to the Hamiltonian in Eq. (\ref{eq:QK_Hamiltonian}) \cite{okamoto1992,eggert1996,sorensen2007quantum,bayat2010negativity,alkurtass2016entanglement}. A negativity probe of this model after a boundary quench may yield information not only about the buildup of the screening cloud, but also about its revival dynamics when confined to a finite geometry.

Another line of future exploration is to find practical applications for the oscillating entanglement cloud. The bond quench indeed results in entanglement distribution between the boundary spin at the quenched side and a few spins on the other side of the chain.  This long-distance entanglement becomes very pronounced when the cloud shrinks to the opposite site of the quenched bond, see the panel II in Fig.~\ref{travelling}(c). Generating such long-distance entanglement has been very appealing~\cite{bayat2010entanglement,alkurtass2013quench,alkurtass2014optimal,banchi2011long,campbell2011propagation,campos2007qubit}. For instance, such entanglement can potentially be used for quantum teleportation~\cite{bennett1993teleporting} and entanglement distillation~\cite{Bennett1996}.

\textbf{Acknowledgments.--} HJ thanks Jean-S\'ebastien Caux for helpful communication. AB acknowledges support from the National Natural Science Foundation of China (Grants No. 12050410253, No. 92065115, and No. 12274059), and the Ministry of Science and Technology of China (Grant No. QNJ2021167001L). SB acknowledges support from UKRI grants EP/R029075/1, EP/N031105/1, EP/S000267/1, and EP/X009467/1 and grant ST/W006227/1.


%

\end{document}